\documentclass[aps,pre,showpacs,twocolumn,groupedaddress,subeqn]{revtex4}
\usepackage{latexsym}
\usepackage{graphicx}
\usepackage{amsmath}
\usepackage{amssymb}
\usepackage{bm}
\usepackage{float}
\usepackage{color}
\def\gtrless{\raise2.5pt\hbox{$>$}\llap{\lower2.5pt\hbox{$<$}}}

\newcommand{\add}[1]{#1}
\newcommand{\rem}[1]{}
\renewcommand{\epsilon}{\varepsilon}

\begin{document}
\title{DYNAMIC GLASS TRANSITION IN TWO DIMENSIONS}
%

\author{M. Bayer$^1$, J. Brader$^1$, F. Ebert$^1$, E. Lange$^1$,
M. Fuchs$^1$, G. Maret$^1$, R. Schilling$^2$}
\email{rschill@uni-mainz.de},\author{M. Sperl$^1$, and J.~P.
Wittmer$^3$}

 \affiliation{$^1$ Fachbereich Physik, Universit\"at
Konstanz, 78457 Konstanz, Germany} \affiliation{$^2$Institut f\"ur
Physik, Johannes Gutenberg-Universit\"at Mainz, Staudinger Weg 7,
D-55099 Mainz, Germany} \affiliation{$^3$Institut Charles Sadron, 6
rue Boussingault, 67083 Strasbourg, France}
\date{\today}


\begin{abstract}
The question about the existence of a structural glass transition in
two dimensions is studied using mode coupling theory (MCT). We
determine the explicit $d$-dependence of the memory functional of
mode coupling for one-component systems. Applied to two dimensions
we solve the MCT equations numerically for monodisperse hard discs.
A dynamic glass transition is found at a critical packing fraction
$\varphi_c^{d=2} \cong 0.697$ which is above $\varphi_c^{d=3} \cong
0.516$ by about 35\%. $\varphi^d_c$ scales approximately with
$\varphi^d_{\rm rcp}$ the value for random close packing, at least for $d$=2,
3.  Quantities characterizing the local, cooperative 'cage motion' do
not differ much for $d=2$ and $d=3$, and we e.~g.~ find the
Lindemann criterion for the localization length at the glass
transition. The final relaxation obeys the superposition principle,
collapsing remarkably well onto a Kohlrausch law. The $d=2$ MCT
results are in qualitative agreement with existing results from MC
and MD simulations. The mean squared displacements measured
experimentally for a quasi-two-dimensional binary system of dipolar
hard spheres can be described satisfactorily by MCT for monodisperse
hard discs over four decades in time provided the \add{experimental} control parameter
$\Gamma$ \add{(which measures the strength of dipolar interactions)} and the packing fraction $\varphi$ are properly related to
each other.
\end{abstract}

\maketitle
\section{INTRODUCTION}

Static and dynamic behavior of macroscopic systems depends
sensitively on the spatial dimension $d$. For example,
one-dimensional systems with short-range interactions do not exhibit
an equilibrium phase transition. \add{In two dimensions there is no long
range order if the ground state exhibits a spontaneously broken continuous symmetry,} and Anderson localization occurs for almost all eigenstates of a disordered
system for $d$=1 and $d$=2, but not in $d$=3, if the disorder is
small. Critical exponents at continuous phase transitions depend on
dimensionality. Concerning dynamical features it is known, for
instance, that the velocity autocorrelation function of a liquid
exhibits a long-time tail proportional to $t^{-d/2}$. Consequently,
the diffusion constant is infinite for $d\leq2$. These few examples
demonstrate the high sensitivity of various physical properties on
the dimension $d$.

Let us consider a liquid in $d$=3. If crystallization can be
by-passed a liquid undergoes a structural glass transition.
Although not all features of this transition are completely
understood, recently significant progress has been made
concerning its microscopic understanding. Following many decades of several phenomenological descriptions with
less predictive power, the mode coupling approach introduced in 1984 by
Bengtzelius, G\"otze and Sj\"olander \cite{1}, has led to a {\it microscopic
theory} of the structural glass transition.

This theory, called mode coupling theory (MCT), has been discussed
theoretically in great detail by G\"otze and his coworkers (see Ref.
\cite{2} for a review). Its numerous predictions were largely
successfully checked by experiments and simulations \cite{3,4}. The
main prediction of MCT is the existence of a {\it dynamical} glass
transition at which the dynamics changes from ergodic to nonergodic
behavior. Thermodynamic (equilibrium) quantities, e.g.~the
isothermal compressibility and structural ones like the static
structure factor $S(q)$, do {\it not} become singular at the glass
transition singularity of MCT. Hence, the MCT glass transition is of
pure dynamical nature. It can be smeared-out by additional
relaxation channels, and then marks a crossover \cite{2,3}.

A {\it microscopic} theory predicting a structural glass transition
with pure thermodynamic origin was derived by M\'ezard and Parisi
\cite{5}. Their replica theory is a first principles approach which
yields a so-called Kauzmann temperature $T_K$ at which the
configurational entropy vanishes. $T_K$ is below $T_c$, the MCT
glass transition temperature. In two dimensions, $T_K$ may mark a
crossover \cite{moore}. For a review of both microscopic theories as
well as phenomenological approaches to the structural glass
transition the reader may consult Ref. \cite{6}.

An important question is now: ``What is the dependence of the
structural glass transition on the spatial dimensionality''? This
question has already been asked by several people some time ago.
Before we come to a short review of this work, let us consider
monodisperse hard spheres and hard discs in $d$=3 and $d$=2,
respectively.

The most dense packing of four hard spheres corresponds to a regular
tetrahedron. However, three-dimensional space cannot be covered
completely by regular tetrahedra, without overlapping. This kind of
geometrical frustration is absent in two dimensions. The densely
packed configuration of three hard discs corresponds to an
equilateral triangle. Since the two-dimensional plane can be tiled
completely without overlap by equilateral triangles, there is no
frustration. Therefore one may be tempted to conclude that there is
no structural glass transition in two dimensions. However, the
link between frustration and glass transition has proven subtle.
Experiments \cite{7} and simulations \cite{8,9} of monodisperse hard
spheres in three dimensions have shown crystallization. Hence, the
existence of geometric frustration is not sufficient for glass
formation. What one needs is bi-dispersity or polydispersity.

Santen and Krauth have performed a MC simulation of a
two-dimensional system of polydisperse {\it hard} discs. The
polydispersity has been quantified by a parameter $\epsilon^{\rm pol}$.
Their results clearly demonstrate (i) the absence of a thermodynamic
glass transition and (ii) the existence of a dynamic glass
transition at a critical packing fraction $\varphi_c(\epsilon^{\rm pol})$.
The kinetic glass transition shifted outside the region of
crystallization for  $\epsilon^{\rm pol}\geq \epsilon^{\rm pol}_{\rm min}\approx
$ 10\% \cite{10,11}. Furthermore, the diffusivity was found to be
consistent with the MCT result \cite{2,3}:
$$D \sim (\varphi_c-\varphi)^\gamma \,\, , \,\, \varphi \leq
\varphi_c$$
where $\gamma \approx$ 2.4 and $\varphi^{\rm sim}_c \approx$0.80.
This critical value agrees with what has been found for a related
system by Doliwa and Heuer (see Figure 2 in Ref. \cite{12}). The
absence of a thermodynamic glass transition has been strengthened
recently by Donev et al.~\cite{13} for a binary hard-disc mixture.

There are a few investigations of glass formation in
two-dimensional systems with {\it soft} potentials. Lan\c{c}on and
Chaudhari \cite{14} studied a binary system with modified Johnson
potential. They found that the structural relaxation time
seems to diverge when approaching a critical temperature. Similar
behavior was observed by Ranganathan \cite{15} for a {\it
monodisperse} Lennard-Jones systems and by Perera and Harrowell
\cite{16} for a {\it binary} mixture of soft discs with a
$1/r^{12}$ potential. It is surprising that the intermediate
self-scattering function $S^{(s)}(q,t)$ of the monodisperse system
\cite{15}, exhibits strong stretching, one of the characteristics
of glassy dynamics. However, $S^{(s)} (q,t)$ does not produce a
well pronounced plateau \cite{15} under an increase of the
density, i.e.~the cage effect does not become strong enough. This
is quite different to the binary system \cite{16}. $S^{(s)} (q,t)$
develops a {\it two-step relaxation} process upon supercooling
with a well pronounced plateau over 4-5 decades in time, at
lower temperatures. This behavior is qualitatively identical to
that found for, e.g.~the Lennard-Jones mixture investigated and
analyzed in the framework of MCT by Kob and Anderson \cite{17}.
The authors of Refs. \cite{10,11,12,13,14,15,16} conclude that
there is a structural glass transition in two dimensions. Their
conclusion is supported by recent experiments on colloidal
particles with repulsive dipolar interactions in two dimensions
\cite{18}. Since these simulational and experimental findings
strongly resemble the MCT predictions obtained for $d$=3, it is
important to apply MCT to two-dimensional liquids. MCT has been
applied to the two-dimensional Lorentz model of overlapping hard
discs \cite{19} and a charged Bose gas with quenched disorder and
logarithmic interactions at zero temperature \cite{20}, but to our
best knowledge not to a two-dimensional liquid-glass problem with
self-generated disorder. To accomplish this is the main motivation
of the present contribution.

The outline of our paper is as follows. The MCT equations for {\it
arbitrary} dimensions and the major predictions of MCT will be
presented in Sec. II. The theory requires as only input the
static structure factor, which is computed in Sec. III. In Sec. IV we apply MCT to a
two-dimensional system of monodisperse hard discs and will
demonstrate that there is a dynamic glass transition. The dynamic
behavior close to that transition is qualitatively identical to
that of monodisperse hard spheres in three dimensions. It will
also be shown that the MCT-result for the time dependent mean
squared displacement describes the experimental result of Ref.
\cite{18} for both species rather satisfactorily over 4 decades in
time. The final section V contains a short summary and some
conclusions.

\section{MODE COUPLING EQUATIONS}
In this section we will shortly review the MCT equations for the
collective and tagged particle correlator of density fluctuations of
a one-component liquid and will present the properties of their
solution close to the glass transition singularity. The only
dependence on dimension $d$ comes through the integrations
element $(2 \pi)^{-d} d^d k \sim (2 \pi)^{-d} k^{d-1} dk$ which
appears in the memory kernels.

MCT  provides equations of motion for the normalized intermediate
scattering function $\phi_q(t)$ and the tagged particle correlator
$\phi^{(s)}_q (t)$. The mathematical {\it structure} of these
equations does {\it not} depend on $d$. For Brownian dynamics
which is appropriate for colloidal systems they read for $d$
arbitrary
\begin{equation} \label{eq1}
\gamma_q \dot{\phi}_q (t) + \phi_q (t) + \int\limits^t_0 dt' m_q
(t-t') \dot{\phi}_q (t') =0 \quad
\end{equation}
with the memory kernel $m_q(t)$ containing fluctuating stresses
and playing the role of a generalized friction coefficient. It
arises because the density fluctuations captured in $\phi_q(t)$ are
affected by all other modes in the system. In MCT, one assumes that
the dominating contributions at long times are given by density pair
fluctuations and approximates (in the thermodynamic limit)
\begin{equation} \label{eq2}
m_q(t) \equiv \mathcal{F}_q [\phi_k (t)] = \int\!\!\!\frac{d^d k}{(2 \pi)^d}\; V ({\bf q}, {\bf k}, {\bf p}) \phi_k (t) \phi_{p} (t)
\end{equation}
The vertices express the overlap of fluctuating stresses
with the pair density modes, and are uniquely determined by the
equilibrium structure
\begin{equation} \label{eq3}
V ({\bf q}, {\bf k}, {\bf p}) =\frac{n}{2} \frac{S_q S_k S_p}{
q^{4}} [{\bf q} \cdot {\bf k} c_k + {\bf q} \cdot {\bf p} c_p]^2\,
\delta({\bf q}-{\bf k}-{\bf p})
\end{equation}
where $n$ is the number of particles per $d$-dimensional volume,
$S_q$ the static structure factor and $c_q$ the direct correlation
function related to $S_q$ by the Ornstein-Zernike equation.
$\gamma_q$ is a characteristic microscopic time scale. The reader
should note that the vertices, Eq.~(\ref{eq3}), have been
approximated by neglecting static three-point correlations.

The
corresponding equations for the tagged particle correlator and $d$
arbitrary are of the same form
\begin{equation} \label{eq4}
\gamma_q^{(s)} \dot{\phi}^{(s)}_q (t) + \phi_q^{(s)} (t) +
\int\limits_0^t dt' m_q^{(s)} (t-t') \dot{\phi}^{(s)}_q (t') =0
\end{equation}
with
\begin{eqnarray} \label{eq5}
m^{(s)}_q (t) &\equiv& \mathcal{F}^{(s)}_q [\phi_k(t), \phi^{(s)}_k
(t) ] \nonumber\\ &=& \int\!\!\! \frac{d^dk}{(2 \pi)^d} \, V^{(s)} ({\bf q}, {\bf k},
{\bf p}) \phi_k (t) \phi^{(s)}_p (t)
\end{eqnarray}
and
\begin{equation} \label{eq6}
V^{(s)} ({\bf q}, {\bf k}, {\bf p}) = n S_k q^{- 4} ({\bf q} \cdot
{\bf k} )^2 (c^{(s)} _k)^2 \delta({\bf q}-{\bf k}-{\bf p}) \quad .
\end{equation}
$c^{(s)}_k= \langle \rho^{(s)*}_{\bf q} \rho _{\bf q} \rangle /(n
S_q)$ is the tagged particle direct correlation function. If the
tagged particle is one of the liquid's particles it is
$c^{(s)}_k=c_k$.

The static correlation functions $S_k, S_p, S_q, c_k, c_p$ and the
correlators $\phi_k(t)$, $\phi^{(s)}_k (t)$, $\phi_p(t)$,
$\phi_p^{(s)} (t)$ depend on $q=|{\bf q}|$, $ k=|{\bf k}|$ and
$p=|{\bf p}|$, only, due to the isotropy of the liquid and glass
phase. Therefore the $d$-dimensional integrals in Eq.~(\ref{eq2})
and Eq.~(\ref{eq5}) can be reduced to a two-fold integral over $k$
and $p$. This will make explicit the $d$-dependence of the vertices.
As a result one obtains, similarly to the case in $d=3$ \cite{1},
\begin{widetext}
\begin{eqnarray} \label{eq7}
\mathcal{F}_q [\phi_k(t)]=n \frac{\Omega_{d-1}}{(4 \pi)^d}
\frac{S_q}{q^{d+2}} \int\limits^\infty_0\!\! dk \int\limits^{q
+k}_{|q-k|}\!\!\!\! dp \, \frac{k\, p\ S_k \, S_p}{[4q^2 k^2 -(q^2+k^2-p^2)^2]^{\frac{3-d}{2}}}\, 
\left[(q^2 +k^2 -p^2)c_k + (q^2-k^2 + p^2)c_p\right]^2 \, \phi_k (t)
\phi_p (t)
\end{eqnarray}
and
\begin{eqnarray} \label{eq8}
\mathcal{F}^{(s)}_q [\phi_k(t) , \phi^{(s)}_k (t)]= 2 n
\frac{\Omega_{d-1}}{(4 \pi)^d} \, \frac{1}{q^{d+2}}
\int\limits^\infty_0\!\! d k \int\limits^{q +k}_{|q-k|}\!\!\!\! dp
\,\frac{ k\, p \;
S_k}{[4 q^2 k^2 -(q^2 + k^2 -p^2)^2]^{\frac{3-d}{2}}} 
\left[(q^2 + k^2 -p^2) c^{(s)}_k\right]^2 \phi_k (t) \phi^{(s)}_p (t)
\end{eqnarray}
\end{widetext}
with
\begin{equation} \label{eq9}
\Omega_d = \frac{2 \pi^{\frac{d}{2}}} {\Gamma(\frac{d}{2})}
\end{equation}
the well-known result for the surface of a $d$-dimensional
unit-sphere. $\Gamma(x)$ is the Gamma function. Note that
Eq.~(\ref{eq9}) yields (with $\Gamma(\frac{1}{2})= \pi^{1/2})$
\begin{equation} \label{eq10}
\Omega_1=2
\end{equation}
which is consistent that the ``one-dimensional unit-sphere'' is an
interval of length two with a ``surface'' consisting of two points.

The behavior for $q \rightarrow 0$ of both functionals can be
obtained by a Taylor expansion of $\int\limits^{k + q}_{k -q} dp
(\cdots)$. Although straightforward it is rather tedious.
Alternatively, one can start directly from
Eqs.~(\ref{eq2}) and ~(\ref{eq5}), in order to obtain
 for $q \rightarrow 0$
\begin{widetext}
\begin{eqnarray} \label{eq11}
\mathcal{F}_q [\phi_k (t)] \to \frac{n}{2} \frac{\Omega_{d}}{2 \pi
^d} S_0 \int\limits_0^\infty dk\, k^{d-1} S_k^2 \left[c_k^2 +
\frac{2}{d} k c_k c'_k + \frac{3}{d(d+2)} k^2 c'^2_k\right]  (\phi_k
(t))^2 + {\cal O}(q)
\end{eqnarray}
and
\begin{equation} \label{eq12}
\mathcal{F}^{(s)}_q [\phi_k (t), \phi^{(s)}_k (t)] \to n
\frac{\Omega_{d}}{d (2 \pi)^d} \, \frac{1} {q^2}
\int\limits_0^\infty dk \, k^{d+1} S_k (c_k^{(s)} )^2 \phi_k (t)
\phi^{(s)}_k (t) + O (1/q) \quad .
\end{equation}
\end{widetext}
Note that $\mathcal{F}_0 [\phi_k (t)]$ exists, whereas
$\mathcal{F}^{(s)}_q [\phi_k(t), \phi^{(s)}_k (t)]$ diverges like
$q^{-2}$. This divergence is related to the absence of momentum
conservation for the tagged particle.

Taking $d$=3 in Eqs. ~(\ref{eq7}),~(\ref{eq8}),~(\ref{eq11}) and
~(\ref{eq12}) one arrives at the well-known representations of the
memory kernel for finite $q$ and $q \rightarrow 0$ \cite{1,21,22}.
 Note that Refs. \cite{21} and \cite{22} already
present the integrals in Eqs.~(\ref{eq7}) and~(\ref{eq8}) in
discretized form.

With the knowledge of the number density, of the static
correlators $S_q$, $c_q$ and $c^{(s)}_q$ as function of the
thermodynamic variables and of $\gamma_q$ and $\gamma^{(s)}_q$ one
can solve Eq.~(\ref{eq1}) and Eq.~(\ref{eq4}) for initial
conditions $\phi_q(0)=1$ and $\phi^{(s)}_q (0)=1$. There exist
several quantities which characterize the solutions. These
quantities can be found in Refs. \cite{2}, \cite{21} and
\cite{22}. In order to keep our presentation self-contained as
much as reasonable we discuss those for which results will be
reported in the next section. We start with the glass transition
singularity. At the  glass transition the nonergodicity parameters
(NEP)
\begin{subequations}
\label{eq13a}
\begin{equation}
 f_q = \lim\limits_{t \rightarrow \infty}
\phi_q(t)
\end{equation}
change discontinuously from zero to a positive nonzero value,
smaller or equal to one. The corresponding quantity
\begin{equation}
\label{eq13b}
 f^{(s)}_q= \lim_{t \rightarrow \infty} \phi^{(s)}_q (t)
\end{equation}
\end{subequations}
can change discontinuously at the same point or in a continuous
fashion at higher densities or lower temperatures. Both NEP fulfill
nonlinear algebraic equations \cite{2}
\begin{equation} \label{eq14}
\frac{f_q}{1-f_q} =\mathcal{F}_q [f_k] \quad , \quad
\frac{f_q^{(s)}}{1-f_q^{(s)}} = \mathcal{F}^{(s)}_q [f_k ,
f^{(s)}_k ] \quad .
\end{equation}
$f_q^c$ and $f_q^{(s)c}$ are the NEP at the critical point, e.g.
at $n=n_c$. Since we will apply MCT to $d$-dimensional hard
spheres with diameter $D$ we use in the following the packing
fraction $\varphi=n \Omega_{d-1} (D/2)^d/d$. Above, but close to
$\varphi=\varphi_c$, i.e.~for $0< \varepsilon
\equiv(\varphi-\varphi_c)/\varphi_c \ll 1$ it is:
\begin{equation} \label{eq15}
f_q=f^c_q +h_q [\sqrt{\sigma / (1-\lambda)} + \sigma (\bar{K}_q+\kappa) /
\sqrt{1-\lambda}]
\end{equation}
with the critical amplitude
\begin{equation} \label{eq16}
h_q=(1-f^c_q)^2 e^c_q \, ,
\end{equation}
the separation parameter
\begin{equation} \label{eq17}
\sigma(\varepsilon) = C \varepsilon + O (\varepsilon^2) \quad ,
\end{equation}
and the so-called exponent parameter, $\lambda$, which obeys $0<
\lambda<1$. The second term on the r.h.s. of Eq.~(\ref{eq15}) is the
leading asymptotic result for $f_q$, and $\bar{K}_q+\kappa$ yields
the next-to-leading order correction. $C,$ $\lambda$, $e_q^c$ and
$\bar{K}_q+\kappa$ follow from $\mathcal{F}_q [f_k]$ and its
derivatives with respect to $f_k$ at $\varphi_c$. Particularly,
$e^c_q$ is the right eigenvector $e_q$ belonging to the largest
eigenvalue $E_{\rm max} (\varphi)$ of the stability matrix
$(\partial \mathcal{F}_q[f_k]/\partial f_k)$ at the critical point.
$E_{\rm max} (\varphi)$ is not degenerate since the stability matrix
is non-negative and irreducible. At the critical point the maximum
eigenvalue becomes one, i.e. $\varphi_c$ can be determined from the
condition $E_{\rm max} (\varphi_c)=1$.

At the critical point $\phi_q(t)$ decays to the plateau value
$f^c_q$, $0<f_q^c <1$. Its time dependence is given by
\begin{equation} \label{eq18}
\phi^c_q (t)=f^c_q + h_q (t/t_0) ^{-a} \{1 + [K_q + \kappa (a)]
(t/t_0)^{-a}\} \, , \, t\gg t_0 \quad .
\end{equation}
$K_q$ (not to be confused with $\bar{K}_q)$ and $\kappa (x)$ are again
determined by $\mathcal{F}_q [f_k]$ and its derivatives at
$\varphi=\varphi_c$. They are a measure of the next-to-leading
order contribution with respect to the leading asymptotic result
\begin{equation} \label{eq19}
\phi^c_q (t)= f_q^c + h_q (t/t_0)^{-a} \quad , \quad t \gg t_0 \,,
\end{equation}
the critical law. This critical decay occurs on a time scale $t$
much larger than a typical microscopic time $t_0$. The exponent
$a$ is determined by $\lambda$, only.

For $\varepsilon <0$ two $\sigma$-dependent, divergent time scales
exist
\begin{subequations}
\begin{equation}
\label{eq20a} t_\sigma = t_0\; |\varepsilon|^{-\frac{1}{2a}} \quad ,
\quad \varepsilon \gtrless\, 0
\end{equation}
and
\begin{equation}
\label{eq20b} t'_\sigma = t_0'\;  |\varepsilon|^{-\gamma} \quad ,
\quad \varepsilon < 0
\end{equation}
\end{subequations}
with $\gamma = \frac{1}{2a} + \frac{1} {2b}$, and $t_0'=
t_0/B^{1/b}$, where $B$ is a constant. The so-called von Schweidler
exponent, $b$, follows from $\lambda$, only. $\phi_q(t)$ exhibits a
two-step relaxation. The relaxation for $t/t_\sigma \ll 1$ to the
critical plateau value $f^c_q$ follows from Eq.~(\ref{eq18}) by
replacing $(t/t_0)$ through $(t/t_\sigma)$ and the decay from that
plateau to zero is initiated by the von Schweidler law for $t_\sigma
\ll t \ll t_\sigma' $
\begin{equation} \label{eq21}
\phi_q(t) =f^c_q -h_q (t/t'_\sigma)^b \{1-[K_q+\kappa (-b)]
(t/t'_\sigma)^b\} \quad .
\end{equation}
$K_q + \kappa (-b)$ determine again the next-to-leading order
contribution.

For $\varepsilon>0$ there is a single relaxation process, only.
$\phi_q(t)$ relaxes for $t/t_\sigma \ll 1$ like for $\varepsilon <
0$, and finally the plateau value $f_q$ is reached by an exponential
long time decay.

For $\varepsilon<0$, the final or $\alpha$-relaxation process
describes the decay of the correlators  from the plateau $f^c_q$
down to zero.  Asymptotically close to the transition, the
functional form of the $\alpha$-process is given by a master
function $\tilde\phi_q(\tilde t)$ of the rescaled time $\tilde t=
t/t_\sigma'$ via
\begin{equation}
\phi_q(t)=\tilde\phi_q(\tilde t) + \epsilon\; \tilde\phi_q^{(2)}(\tilde t) +  {\cal O}(\epsilon^2) \quad\mbox{for }\, \epsilon \to 0-
\label{eq21z}
\end{equation}
The master function $\tilde\phi_q$ obeys an equation similar to
Eq.~(\ref{eq1}) with vertices evaluated right at the critical point,
$\epsilon=0$. Thus it does not depend on separation $\epsilon$ and
control parameters, and Eq.~(\ref{eq21z}) expresses the often
observed '(time-temperature) superposition principle' \cite{2,3}.
The von Schweidler series  Eq.~(\ref{eq21}) gives the short time
behavior of $\tilde\phi_q$ for $\tilde t\to0$, and the corresponding
result for the correction is:  $\tilde\phi_q^{(2)}(\tilde t\to0) \to
h_q B_1 C \,\tilde t^{-b}$, with $C$ from Eq.~(\ref{eq17}) and $B_1$
a known constant.

 Similar leading order and next-to-leading
order contributions can be derived for the tagged particle
correlator $\phi^{(s)}_q (t)$, and e.g.~the mean squared
displacement $\delta r^2 (t)= \langle ({\bf r}(t)-{\bf r}(0))^2
\rangle$ \cite{22}. Since
\begin{equation} \label{eq22}
\delta r^2 (t) = \lim\limits_{q \to 0} \frac{2d}{q^2}
[1-\phi^{(s)}_q (t) ] \quad ,
\end{equation}
the long-wave limit of Eq.~(\ref{eq4}) yields after integration
with respect to $t$:
\begin{equation} \label{eq23}
\delta r^2(t) + D^{(s)}_0 \int^t_0 dt' \widetilde{m}_0^{(s)}
(t-t') \delta r^2 (t') = 2 d D_0^{(s)} t
\end{equation}
where the memory kernel $\widetilde{m}_0^{(s)} (t)$ follows from
Eqs.~(\ref{eq5}) and Eq.~(\ref{eq12})
\begin{eqnarray} \label{eq24}
\widetilde{m}^{(s)}_0 (t) &\equiv& \lim\limits_{q \rightarrow 0} q^2
\mathcal{F}^{(s)}_q [\phi_k (t), \, \phi_k^{(s)} (t)]\nonumber\\
&=& n \frac{\Omega_{d-1}}{d(2 \pi)^d} \, \int\limits_0^\infty dk \,
k^{d +1} S_k (c_k^{(s)})^2 \phi_k (t) \phi^{(s)}_k (t) \quad .
\end{eqnarray}
Furthermore we have used:
\begin{equation} \label{eq25}
\gamma^{(s)}_q = 1 / (D^{(s)}_0 q^2)
\end{equation}
with $D^{(s)}_0$ the short-time diffusion constant of the tagged
particle.
Eqs.~(\ref{eq13b}) and Eq.~(\ref{eq22}) imply that the long-time
limit of $\delta r^2 (t)$
\begin{equation} \label{eq26}
\lim\limits_{t \rightarrow \infty} \delta r^2 (t)= 2d r^2_s
\end{equation}
is related to the tagged particle's localization length $r_s$
given by:
\begin{equation} \label{eq27}
r^2_s = \lim\limits_{q \rightarrow 0} \frac{1-f_q^{(s)}}{q^2}
\quad .
\end{equation}
In the liquid phase where $f^{(s)}_q=0$, Eq.~ (\ref{eq26}) gives $r_s=\infty$,
i.e.~the particle is delocalized. While $r_s$ becomes finite at the
glass transition.

Besides the correlators $\phi_q(t)$ and $\phi^{(s)}_q (t)$ one can
also study the corresponding susceptibilities $\chi_q(\omega)$ and
$\chi^{(s)}_q (\omega)$, respectively. Similar asymptotic laws and
next-to-leading order correction exist for them \cite{21,22}.
Independent on whether the correlators or their susceptibilities
are considered the dependence on $d$ of the leading and
next-to-leading order terms enters only via the $d$-dependence of
$\mathcal{F}_q$ (Eq.~(\ref{eq7})) and $\mathcal{F}^{(s)}_q$
(Eq.~(\ref{eq8})).

\section{STATIC STRUCTURE}

In this section we consider the calculation of accurate equilibrium
structural correlation functions for the hard disc system. Within
MCT, all information regarding the interparticle interactions is
contained in the static structure factor which enters the memory
function vertices Eqs.~(\ref{eq3}) and (\ref{eq6}); the interaction
potential does not enter explicitly in the MCT equations. Experience
with MCT calculations in three dimensional systems has shown that
the location of the glass transition somewhat depends on the details
of the input structure factor, particularly the height of the main
peak. Different approximate theories for the static structure lead
to varying values for e.~g.~the critical packing fraction
$\phi_c^{(d=3)}$ \cite{Barrat}. We have therefore considered a
number of approximation schemes for the two dimensional static
structure factor in order to obtain the best possible values for the
description of the ideal glass transition. The quality of the
various approximation schemes is assessed by comparison with
computer simulation data. \add{(Monte Carlo simulations of $6\, 10^4$
particles,  as well as event
driven molecular dynamics simulations of 1089 particles were performed, both in the NVT
ensemble \cite{allen}.)}

Integral equation theories based on the Ornstein-Zernike (OZ)
equation provide a powerful method to calculate the pair correlation
functions for a given interaction potential \cite{hansen}. The OZ
equation is given by
\begin{equation} \label{oz}
h(r) = c(r) + n\int d^{d}r' \;c(r')\,h(r-r'),
\end{equation}
where $h(r)\equiv g(r)-1$.
This expression must be supplemented by an additional (generally approximate)
closure relation between $c(r)$ and $h(r)$.
For hard spheres in $d$-dimensions the most widely used closure is the
Percus-Yevick relation $g(r<1)=0$, $c(r>1)=0$.
In odd dimensions the resulting integral equation can be solved analytically
for the direct correlation function $c(r)$.
In even dimensions there exists no analytic solution and full numerical solution
is required \cite{lado}.
Efforts have been made to approximate the numerical PY data by analytic forms
\cite{baus-colot,rosenfeld} but in all cases these fail to reproduce accurately  the
detailed structure of the numerical solution at high densities.
The formally exact closure to the OZ equation for systems with pairwise interactions
is given by
\begin{equation} \label{bridge}
h(r)=-1 +\exp(-\beta u(r) + h(r)-c(r) +B(r)),
\end{equation}
where $u(r)$ is the pair potential and $B(r)$ is the bridge function, an intractable
function representing the sum of the most highly connected diagrams in the
virial expansion. Setting $B(r)=0$ recovers the familiar hypernetted-chain
approximation.
\begin{figure}[H]
\begin{center}
\includegraphics[width=\columnwidth]{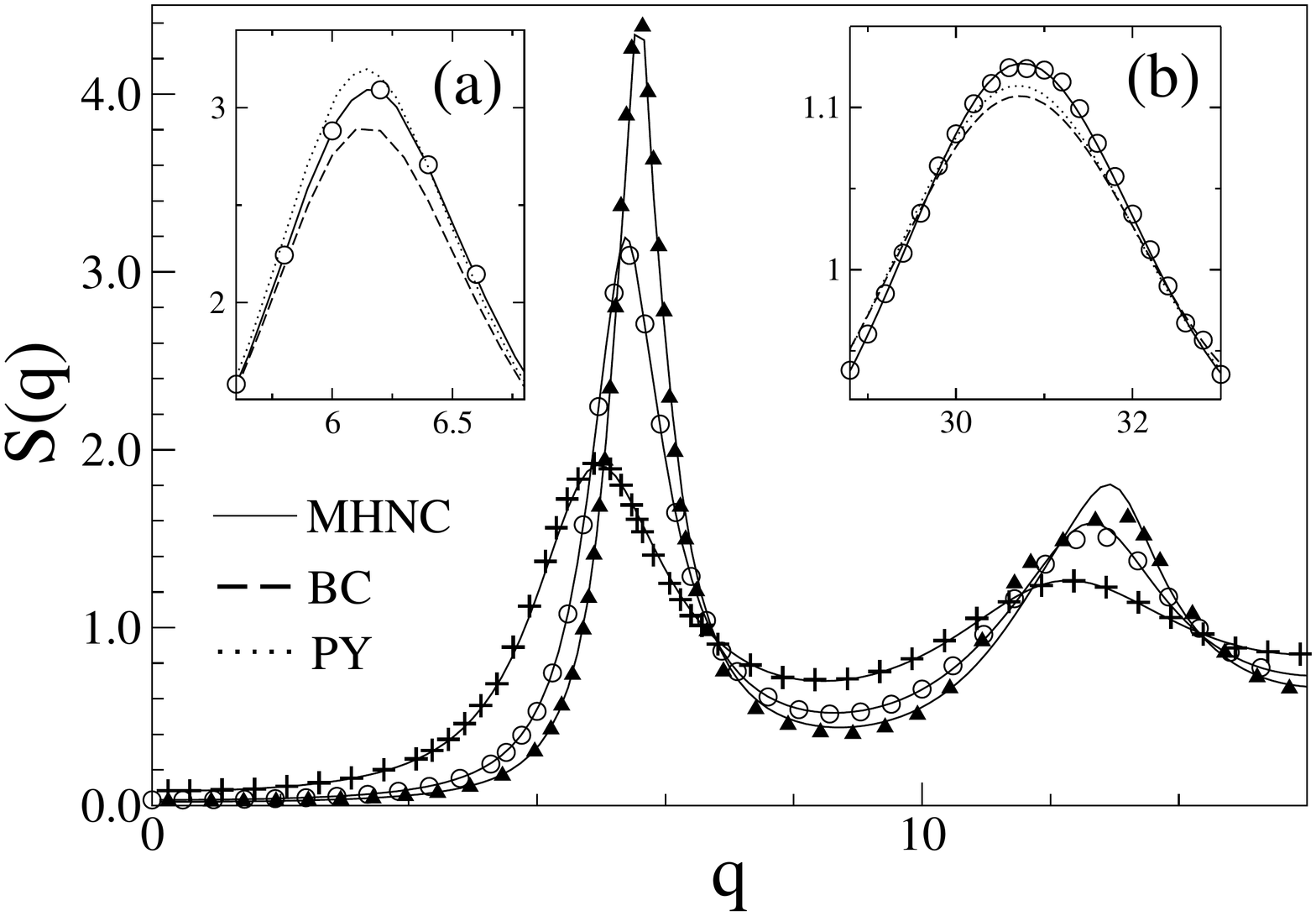}
\caption{Comparison between theoretical MHNC structure factors and
simulation for hard discs at packing fractions $\phi=0.5$ ($+$), $0.628$ ($\circ$),
and $0.68$ ($\triangle$); $q$ is given in units of  $1/D$, the inverse diameter.
Inset (a) concentrates on the vicinity of the main peak and gives
additional comparison with Baus-Colot and PY theories for packing
fraction $\phi=0.628$. Inset (b) demonstrates how the MHNC theory
correctly captures the asymptotic behaviour for large $q$ values for
packing fraction $\phi=0.628$.} \label{statik}
\end{center}
\end{figure}

The modified-hypernetted-chain (MHNC) approximation is to take
$B(r)$ from the PY theory solved at some {\em effective} density
$n^*$, different from the true system density $n$, and to treat this
as a variational parameter to ensure thermodynamic consistency
between the virial and compressibility routes to the pressure. A
detailed description of the MHNC equation can be found in
\cite{mhnc}. The steps taken in solving the MHNC equation are, (i)
numerically solve the PY equation at density $n^*$, (ii) use
Eq.(\ref{bridge}) to find $B(r)\equiv B_{PY}(r;n^*)$, (iii) solve
Eqs.(\ref{oz},\ref{bridge}) with this bridge function, (iv)
calculate the pressure from the virial and compressibility
equations, (v) adjust $n^*$ until the two pressures are equal. In
three dimensions it is generally recognized that the MHNC
approximation provides a highly accurate description of the pair
correlations for the hard sphere fluid, significantly improving upon
the PY theory. We find that the same is true in the case of two
dimensional hard discs. At low densities the MHNC $S_{q}$ lies very
close to the PY result. As the density increases discrepancies begin
to arise, particularly in the region of the main peak, with the MHNC
in closer agreement with simulation. Both the MHNC and PY theories
are significantly more accurate than the analytical Baus-Colot
expression \cite{baus-colot}. Figure \ref{statik} shows a comparison
between the MHNC $S_q$ and the simulation results. The level of
agreement is very satisfactory and the MHNC shows clear improvement
over the other theories investigated. The only notable deviation
from the simulation results is the height and width of the second
(and third) peak, which is overestimated (respectively
underestimated) by the MHNC theory. It is known that upon
approaching the crystallization phase boundary (located at
$\phi_F=0.69$ for hard discs) a shoulder develops on the second peak
of the structure factor; a feature which has been interpreted as an
indicator of approaching crystallization \cite{truskett}. The
development of the shoulder also suppresses the height of the second
peak to some extent and leads to a small shift in the location of
the third peak. The MHNC theory, like the PY and all other standard
integral equation theories do not contain information about
crystallization and thus predict fluid-like structure at, and
beyond, the freezing transition. While this property leads to some
discrepancy with simulation results at high density, it makes such
theories ideal for calculating the fluid-like structure factors
required as input to the MCT, where we assume crystallization to
have been suppressed. We can thus proceed with confidence using MHNC
structure factors as input to the MCT.

\section{RESULTS}

\begin{figure}[H]
\begin{center}
\includegraphics[width=\columnwidth]{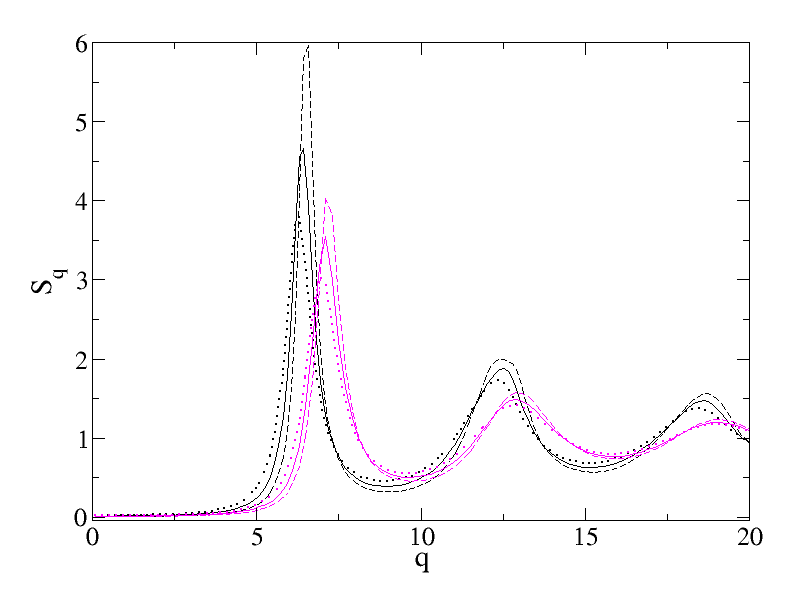}
\caption{\add{(Color online)} Comparison of 2d MHNC structure factors (black/dark) to 3d
PY (magenta/light). The packing fractions correspond to $\epsilon=0$
(solid lines), $\epsilon=-10^{-4/3}$ (dotted lines) and
$\epsilon=+10^{-4/3}$ (dashed lines). The critical packing fractions
are 0.697 for 2d and
0.516 for 3d.} \label{f1}
\end{center}
\end{figure}

In this chapter we will apply MCT to a {\it two-dimensional} system
of monodisperse hard discs with diameter $D$. We will solve the MCT
equations ~(\ref{eq1}),~(\ref{eq4}) and ~(\ref{eq23}) and will
present results for those quantities discussed in the last section.
As input we will use the static structure factors $S_q$ obtained
from the MHNC approach.  This result is presented in Figure \ref{f1}
for three different packing fractions close to the glass transition
and compared with the corresponding Percus-Yevick result for hard
spheres in $d$=3. For instance, for $\varphi=\varphi_c^{d=2}$ and
$\varphi=\varphi_c^{d=3}$, i.e.~for $\varepsilon=0$, the peaks are
more pronounced in $d=2$. In particular the main peak is more narrow
and higher for $d=2$ than for $d=3$. The direct correlation function
$c_q$ follows from the Ornstein-Zernike equation. Because we choose
the tagged particle as one of the liquid particles we have
$c^{(s)}_q=c_q$.

For the friction coefficients in Eqs.~(\ref{eq1}) and~(\ref{eq4}) we
take $\gamma_q=S_q/(D^{(s)}_0 q^2)$ and $\gamma^{(s)}_q =
1/(D^{(s)}_0 q^2)$ (cf.~Eq.~(\ref{eq25})), and choose as our unit of
time $\tau^{(BD)}=D^{(s)}_0/(10D^2)$; in the following all times
will be given as rescaled ones, $t/\tau^{(BD)}$. For the numerical
solution of MCT equation one has to discretize $q$. A compromise
between a fine grid and computation time is required, as the
computations scale with number of gridpoints $M^3$. We choose a grid
with $M=250$ gridpoints and a high q cutoff of $50/D$. A higher
cutoff has only a very small effect on the critcal packing fraction
($<10^{-4}$). The effect of the number of gridpoints is more
sensitive due to the form of the Jacobian of the transformation to
bipolar coordinates. In this paper the integration is substituted by
a rule that can be called a modified trapezoid rule. The value of
the function to be integrated is not taken in the middle of the
interval $[n h,(n+1) h]$ but at $(n + 0.303)h$. By using this rule
on gets the best discrete description of the Jacobian hence it is
used here. The difference in $\varphi$ between $M=500$ and $M=250$
is then $<10^{-3}$ leading to a system close enough to continuum.
Then the solution of Eqs.~(\ref{eq1}),~(\ref{eq4}) and~(\ref{eq23})
will be performed by use of a decimation technique \cite{23}.

\begin{figure}[H]
\begin{center}
\includegraphics[width=\columnwidth]{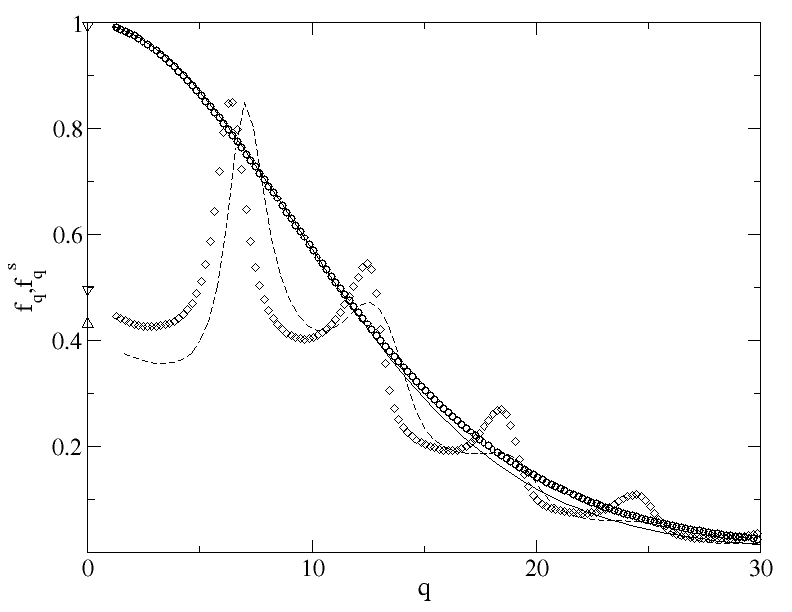}
\caption{Non ergodicity parameter of coherent (2d diamond, 3d dashed
line) and incoherent (2d circles, 3d solid line) correlators at
critical packing fraction $\varphi_c^{2d}=0.697$ and
$\varphi_c^{3d}=0.516$. The q-values for $q<1$ are not included
since they can not be determined accurately for numerical reasons.
The q=0 results are from the analytic expansions Eqs.~(\ref{eq11})
and (\ref{eq12}), and are included as triangles.} \label{Nep}
\end{center}
\end{figure}

The search for the glass transition singularity can be done either
by an iterative solution of the non-linear equations ~(\ref{eq14})
or by calculation of $E_{\rm max} (\varphi)$. In this paper a simple
bracketing algorithm is used starting from two points were point A
yields a finite NEP for q near the peak position and point B has
$f_q=0$. The next point C is taken in the middle of the interval. If
it yields finite $f_q$ the next point is taken between A and C, if
$f_q=0$ the other interval [C,B] is taken. This procedure is
continued until the critical packing fraction $\varphi_c$ is
determined to a precision of $10^{-10}$. As a result we have found
$$\varphi^{d=2}_c \cong 0.696810890 (317)$$
which is above the value for $d =3$ \cite{21}
$$\varphi^{d=3} _c \cong 0.51591213(1) \quad .$$
\add{(The denoted accuracy will be required to reliably
compute $\varepsilon$ in the following.)}
Similar to the three-dimensional system, the collective and self
part of the density fluctuations become nonergodic at the same
critical packing fraction $\varphi^{d=2}_c$. The corresponding
critical nonergodicity parameters are shown in Figure \ref{Nep}.

While almost no difference between incoherent NEPs for $d=2$
and $d=3$ can be observed,  more pronounced maxima appear in the
coherent NEP at higher wavevectors for the lower dimension. Regions
of rather abrupt $q$-dependences in $f_q$ should be observable
experimentally. Two length scales appear to be involved in $f_q$. While the average particle distance, connected to the main peak in $S_q$, somewhat differs from $d=3$ to $d=2$, the localisation length, which dominates the incoherent NEP, is insensitive to dimensionality. The change of $f_q$ when stepping down in dimension thus can not simply be scaled away.

An important observation in the numerical solution of Eq.~(\ref{eq14}) concerns
the convergence of the required integrals. We find here, and for all other integrations performed, that convergence at small and large wavevectors holds. We interpret this as indication that the MCT glass transition in $d=2$ describes a local phenomenon not affected by long range correlations, which might sensitively depend on dimensionality.

\begin{figure}[H]
\begin{center}
\includegraphics[width=\columnwidth]{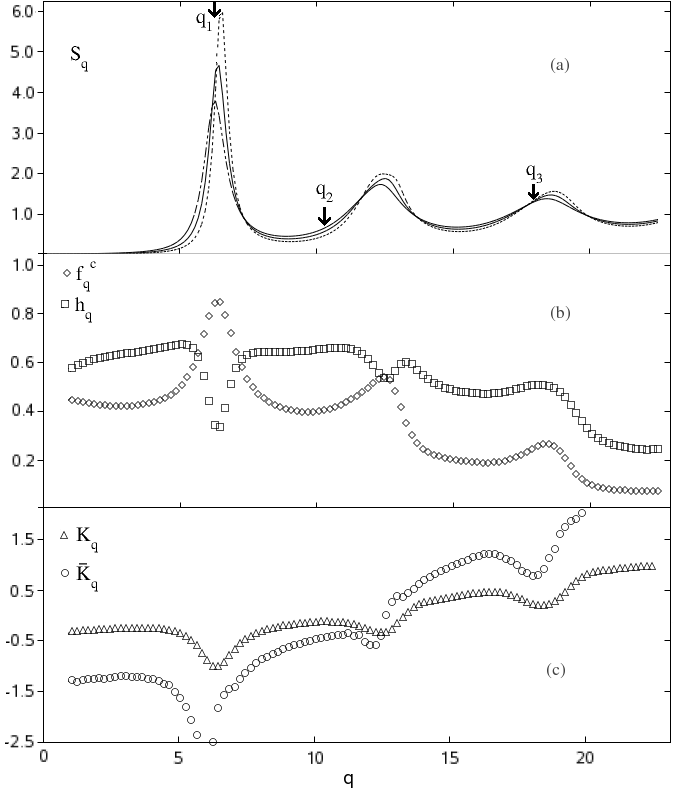}
\caption{(a) Structure factor $S_q$ as function of wave vector q for
$\varphi=\varphi_c\approx0.697$ (solid), $\varphi\approx0.729$
(dotted) and $\varphi\approx0.664$ (dashed); the latter correspond to $\epsilon=\pm10^{-4/3}$.
 The arrows mark the wave vectors $q_1=6.46$,
$q_2=10.06$ and $q_3=18.26$. (b) Critical NEP $f_q^c$ (diamonds) and
critical amplitude $h_q$ (squares). (c) The amplitudes $K_q$
(triangles) and $\bar K_q$ (circles).} \label{Kamplitudes}
\end{center}
\end{figure}

All wavevector dependent structure functions describing the glassy
structure and its relaxation ('cage effect') are summarized in
Fig.~\ref{Kamplitudes}.  The critical nonergodicity parameter
$f_q^c$ and the critical amplitude $h_q$ is depicted in Figure 4b
and the next-to-leading order amplitudes $K_q$ and $\bar{K}_q$
(cf.~Eqs.~(\ref{eq15}), ~(\ref{eq18}) and ~(\ref{eq21})) in Figure
\ref{Kamplitudes}c. The $q$-dependence of the shown quantities
follows generally the one of $S_q$. Only $h_q$ exhibits the opposite
variation around the main structure factor peak. Figure
\ref{Kamplitudes}a specifies three typical wave numbers for which
results will be discussed below. Comparison with the $d=3$ result
(Figure 2 in Ref. \cite{21}) reveals qualitative similar
$q$-dependence of $f^c_q$, $h_q$, $K_q$ and $\bar{K}_q$. Like for
$S_q$, the $q$-variation of all quantities is more pronounced in
$d=2$ than in $d=3$.

\begin{figure}[H]
\begin{center}
\includegraphics[width=\columnwidth]{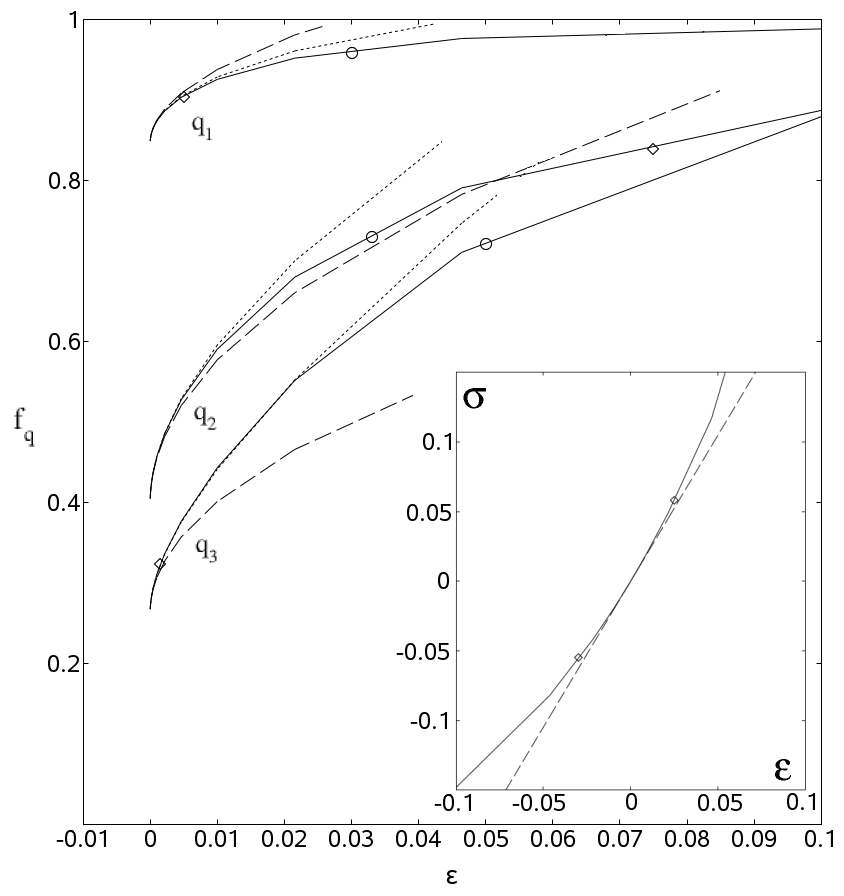}
\caption{Non ergodicity parameter $f_q$ for $q_1=6.46$, $q_2=10.06$
and $q_3=18.26$ (solid lines). The leading asymptotes (dashed)
describe $f_q-f_q^c$ within 10\% up to the values marked by diamonds
(0.005, 0.075, 0.0015). The next to leading asymptotes (dotted) are
within 10\% for $\epsilon$ up to the values marked by circles (0.03,
0.033, 0.05). The values for the correction amplitudes are $C=2.08$,
$\kappa=1.18$, $h_1=0.337$, $h_2=0.654$, $h_3=0.508$, $\bar
K_1=-1.86$, $\bar K_2=-0.456$ and $\bar K_3=0.844$. The inset shows
the separation parameter $\sigma$ as a function of
 $\epsilon$. The dashed line is the linear asymptote $\sigma=C
 \epsilon$ The range where the asymptote deviates less than 10\%
 from $\sigma$ is between the diamonds that are at $\varphi=0.676$
and $\varphi=0.714$.}\label{fig5}
\end{center}
\end{figure}

 As mentioned in section II., the separation parameter
$\sigma(\varepsilon)$ can be calculated from $\mathcal{F}_q[f_k]$
and its derivatives at $\varphi_c$. The result is given in the inset
of Figure \ref{fig5}. The linear term in Eq.~(\ref{eq17}) describes
$\sigma(\varepsilon)$ for $-0.030 \leq \varepsilon \leq 0.025$ with
an accuracy better than 10\%. This range is similar to that for the
corresponding result for $d=3$ \cite{21}, and provides an estimate
for the range of validity of the asymptotic expansions. The quality
of the leading order result and the next-to-leading order
contribution for $f_q$ (cf.~Eq.~(\ref{eq15})) is demonstrated in
Figure 5 for the three $q$-values $q_1$, $q_2$ and $q_3$ (Figure
4a). It is interesting, that for $q_2$ the leading asymptote
describes $f_q(\varepsilon)$ best and for an unexpectedly wide
range. This arises because of a cancellation of higher order
correction terms. For $q_1$ and $q_3$ the next-to-leading order has
to be taken into account already for rather small $\varepsilon$, as
expected from the $\sigma(\epsilon)$-curve. The overall behavior
also is quite similar to $d=3$ (cf.~Figure 3 of Ref.~\cite{21}).

Now we turn to dynamical features. Figure 6 presents the normalized
correlators $\phi_i(t) \equiv\phi_{q_i}(t)$ for $i=1,2$. The
two-step relaxation process for $\varepsilon <0$ becomes obvious for
both correlators. Since $f^c_{q_2} < f^c_{q_1}$ (cf. Figure 4a,b)
the plateau heights for $\phi_2(t)$ are below those for $\phi_1(t)$.
Again, the $t$- and $\omega$-dependence (the latter is not shown) is
in qualitative agreement with the corresponding results in $d=3$
(cf.~ Figures 4 and 6 of Ref. \cite{21}). In the following, we will
apply the asymptotic expansions from Sect.~II to the correlators in
order to characterize the long-time dynamics in more detail.
\begin{figure}[H]
\begin{center}
\includegraphics[width=\columnwidth]{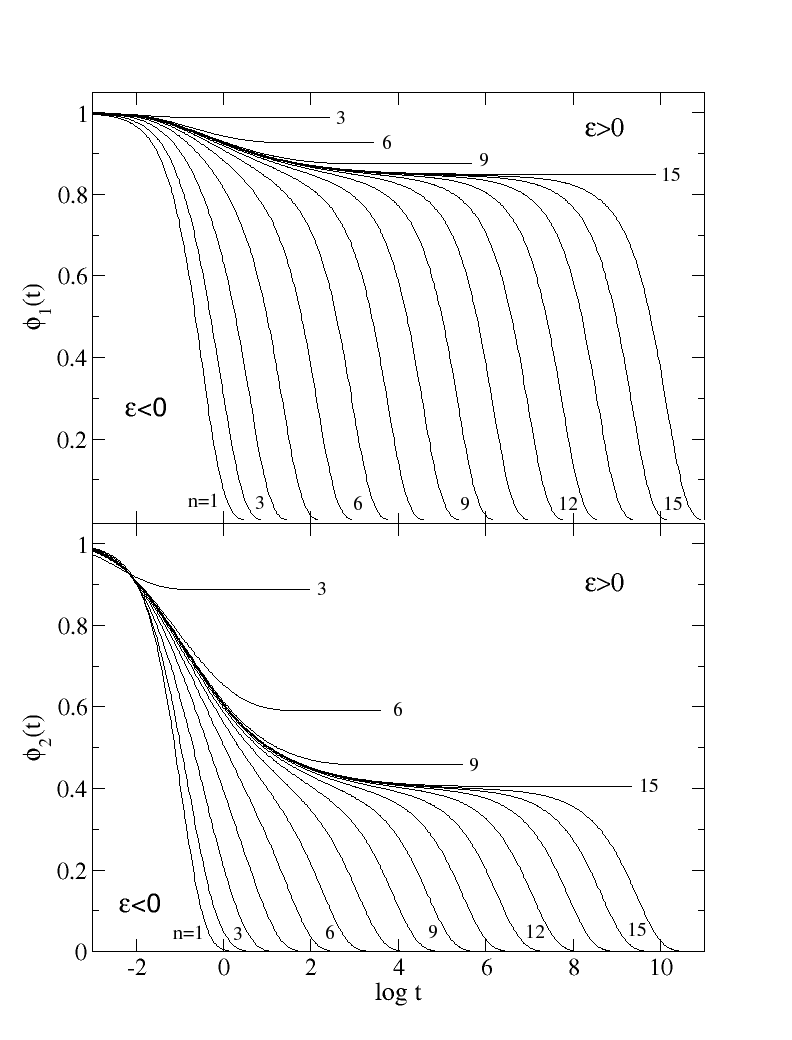}
\caption{Coherent correlators $\phi_1(t)$ (top) and $\phi_2(t)$
(bottom) for different $\epsilon=\pm 10^{-n/3}$. Curves are labeled
by $n$.} \label{correlator}
\end{center}
\end{figure}
Following Ref. \cite{21} we have calculated typical time scales
$\tau^\pm_q (\varepsilon \gtrless 0) $ and $\tau'_q \,
(\varepsilon<0)$ characterizing the first and second relaxation
step. In the fluid, $\tau^-(q)$ marks the crossing of the plateau,
$\phi_q(\tau^-(q))=f^c_q$, and the $\alpha$-relaxation time is
defined by $ \phi_q(\tau'_q)=f^c_q /e$. In the glass, $\tau^+(q)$
captures the approach to the long-time plateau, $
\phi_q(\tau^+_q)-f^c_q=1.001(f_q-f^c_q)$. Results are shown in
Figure \ref{timescale} for $q=q_1$. The divergence of these
relaxation times at $\varepsilon=0$ is described by the asymptotic
laws, Eqs.~(\ref{eq20a}) and~(\ref{eq20b}). Since $\varphi^{d=2}_c$
has been determined, one can calculate the exponent parameter
$\lambda$. As a result we find $\lambda^{d=2} \cong 0.7167$ which
implies $a^{d=2}\cong 0.320$, $b^{d=2} \cong 0.613$ and
$\gamma^{d=2} \cong 2.38$. These values are close to $\lambda
^{d=3} \cong 0.735$, $a^{d=3} \cong 0.312$, $b^{d=3} \cong 0.583$
and $\gamma ^{d=3} \cong 2.46$ \cite{21}. Using in
Eqs.~(\ref{eq20a}) and ~(\ref{eq20b}) $a^{d=2}$ and $\gamma^{d=2}$
leads to the asymptotes (solid lines) in Figure \ref{timescale}.
\begin{figure}[H]
\begin{center}
\includegraphics[width=\columnwidth]{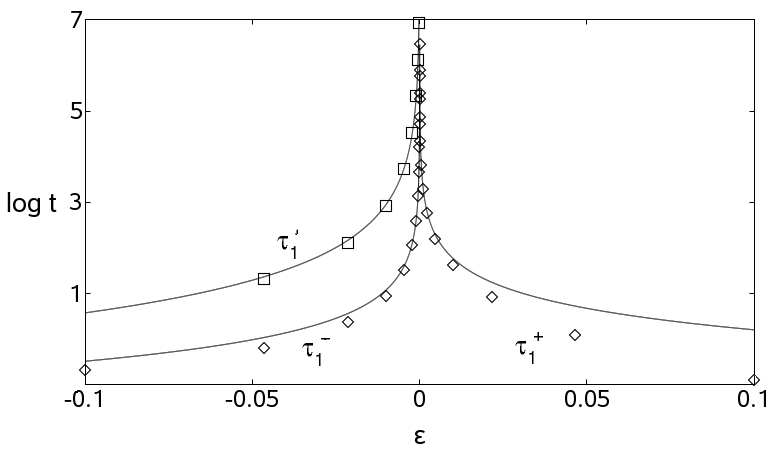}
\caption{Timescales $\tau^{\pm}$ for the $\beta$ process and
$\tau^\prime$ for the $\alpha$ process in the liquid for wave vector
$q_1$. The lines are the asymptotic power laws. $\lambda=0.7167$,
$a=0.320$, $b=0.613$, $\delta=\frac{1}{2a}=1.56$, $c_+=0.044$,
$c_-=0.009$, $\gamma=\frac{1}{2a}+\frac{1}{2b}=2.38$,
$c^\prime=0.0173$ (For the definitions of $c_\pm$ and $c'$ see
Ref.~\cite{21}.)} \label{timescale}
\end{center}
\end{figure}

\begin{figure}[H]
\begin{center}
\includegraphics[width=\columnwidth]{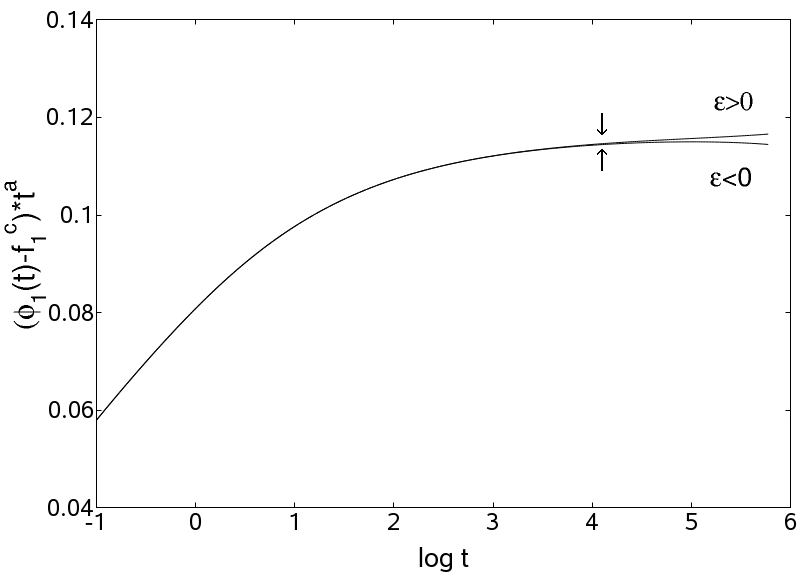}
\caption{Determination of $t_0$. $(\phi_1(t)-f_1^c)t^a$ is plotted
over $\log t$, for  glass and liquid curves close to the transition.
The position where the plateau is best reached by both curves is
marked by arrows. The value found is 0.114. $t_0$ is then given by
$(\dfrac{0.114}{h_1})^{1/a}=0.034$, with $h_1=0.337$ and $a=0.320$.}
\label{t0}
\end{center}
\end{figure}

\begin{figure}[H]
\begin{center}
\includegraphics[width=\columnwidth]{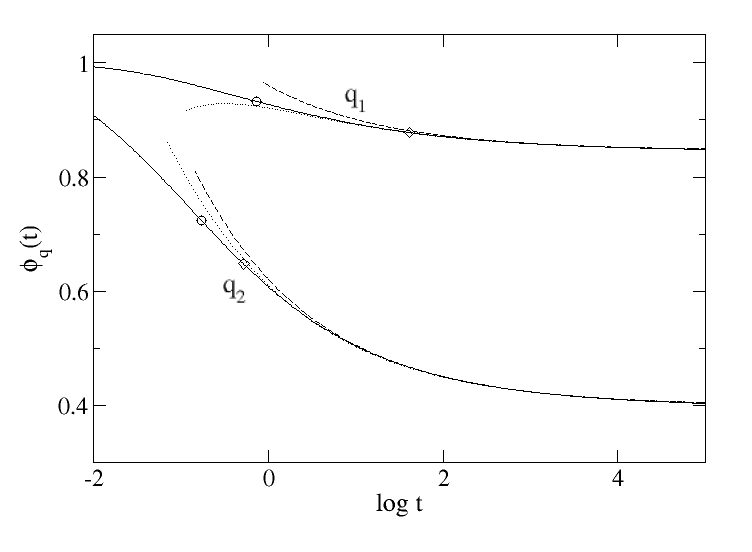}
\caption{The critical laws. The leading asymptotes (dashed) describe
the solution within 10 \% up to the points marked by diamonds (41,
0.52) for ($q_1$, $q_2$). The next to leading results (dotted) are
within 10 \% up to the circles (0.73, 0.17) for ($q_1$, $q_2$).
Correction amplitudes are $\kappa(a)=-0.021$, $K_1=-1.007$,
$K_2=-0.120$.} \label{power}
\end{center}
\end{figure}

\begin{figure}[H]
\begin{center}
\includegraphics[width=\columnwidth]{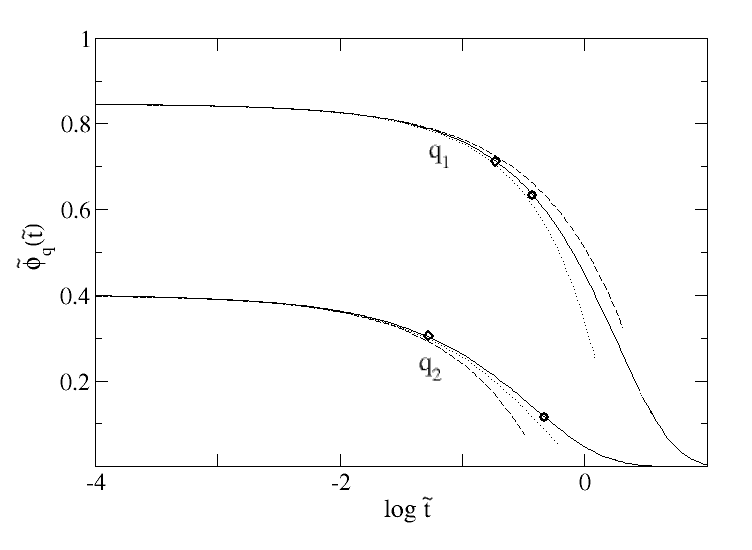}
\caption{The von Schweidler law. The leading asymptotes (dashed)
describe the solution within 10 \% up to the points marked by
diamonds (0.29, 0.053) for ($q_1$, $q_2$). The next to leading
results (dotted) are within 10 \% up to the circles (0.37, 0.47) for
($q_1$, $q_2$) ($b=0.613$ and $t_\sigma^\prime=7.45\, 10^9$).
Correction amplitudes are $\kappa(b)=0.496$, $K_1=-1.007$, $K_2=-0.120$.} \label{schweidler}
\end{center}
\end{figure}

The microscopic time scale $t_0$ entering the critical power law,
Eq.~(\ref{eq18}), can be deduced by plotting $(\phi_q(t)-f_q^c)t^a$
versus log $t$ for $\varepsilon > 0$ and $\varepsilon < 0$ (see Figure
\ref{t0}). The value at which a constant plateau is best reached by
both curves is 0.114=$h_q t_0^a$. With $h_{q_1} \cong 0.337$ and
$a^{d=2}$ we get $t_0=0.034$.
The quality of the leading order result (Eq.~(\ref{eq19})) and its
next-to-leading order correction (second and third term in the curly
bracket of Eq.~(\ref{eq18})) of the critical law is checked in
Figure \ref{power}. The similar check for the von Schweidler law
(Eq.~(\ref{eq21})) is done in Figure \ref{schweidler}. Like for
$f^c_q$, the leading order has a large range of validity for
$q=q_2$.
Note, that both time scales of the structural relaxation are
given by the matching time $t_0$ determined in Fig.~\ref{t0}, and by
the separation parameter $\epsilon$. Thus,  there remains no
adjustable parameter in the test of the von Schweidler law in
Fig.~\ref{schweidler}.

\begin{figure}[H]
\begin{center}
\includegraphics[width=0.9\columnwidth]{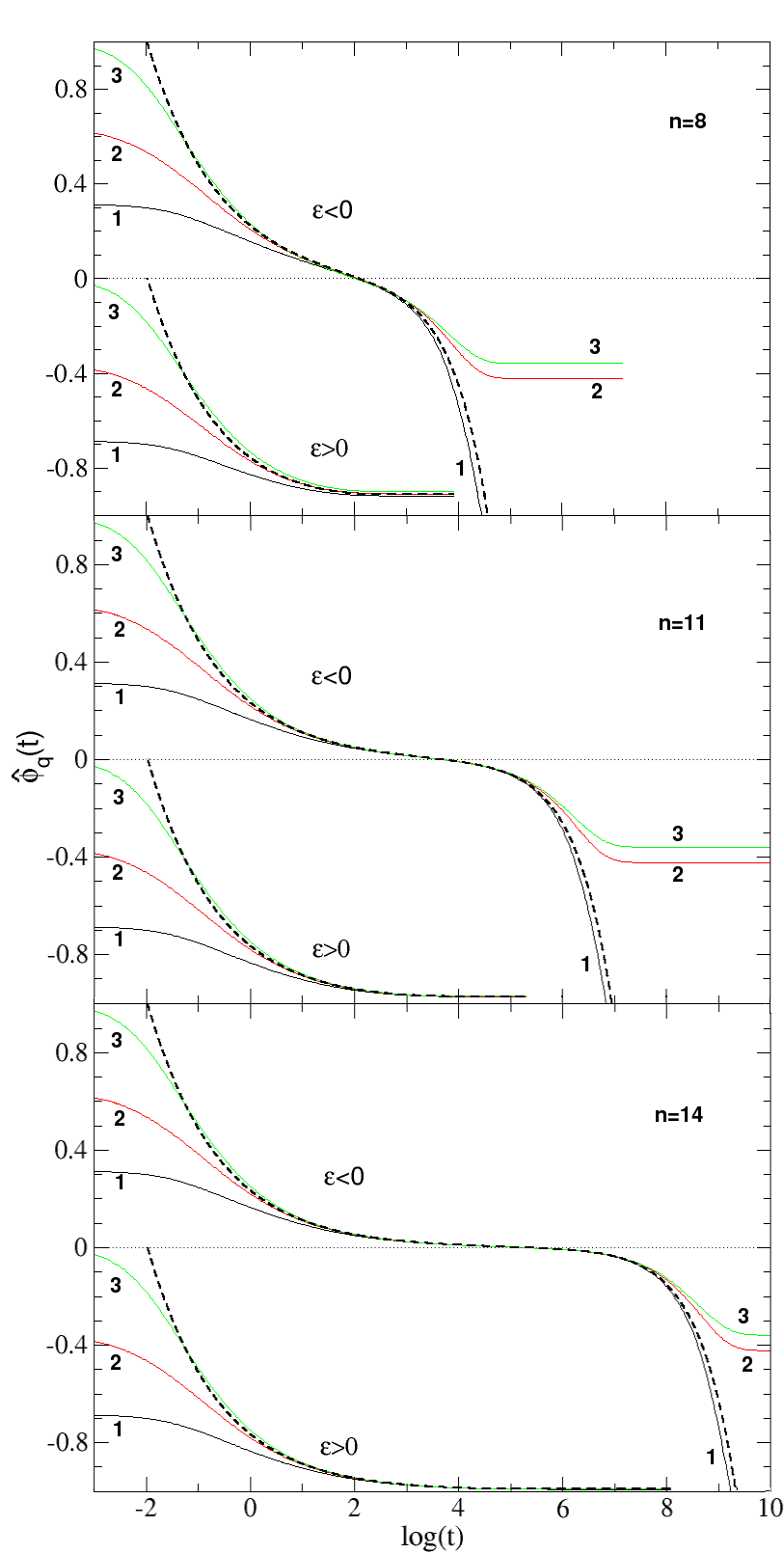}
\caption{\add{(Color online)} Functions  $\hat \phi_q(t)=(\phi_q(t)-f_q^c)/(h_q\sqrt{C})$
for $q_1=6.46$, $q_2=10.06$ and $q_3=18.26$ for $\epsilon=\pm
10^{-n/3}$ for 3 values of n (solid lines). The scaling asymptotes
$G(t)/\sqrt{C}$ are shown as dashed lines. The region where the
functions for different q collapse onto a master function
($\beta$-region) increases with decreasing $\epsilon$. $C=2.08$;
the curves for $\epsilon>0$ are shifted down by 1.} \label{beta}
\end{center}
\end{figure}

The critical law and the von Schweidler law are the short and long
time expansion of the so-called $\beta$-master function $g_\pm
(\hat{t}=t/t_\sigma)$ for $\varepsilon \gtrless \,0$, respectively.
$g_\pm (\hat{t})$ describes the first scaling law regime. The $t$-
and $\varepsilon$-dependence of $\hat \phi_q(t) =
(\phi_q(t)-f^c_q)/(h_q C)$ on the time scale $t_\sigma$ is given by
$g_\pm (\hat{t})$ for $|\varepsilon| \rightarrow 0$, independent on
$q$ \cite{2}. For the two-dimensional system this property is
demonstrated by Figure \ref{beta}. 
We clearly observe that the
curves for $q_1$, $q_2$ and $q_3$ collapse onto a master function
with increasing $n$, i.e.~ for $\varepsilon \rightarrow 0$. Because
of the connection between the $q$-dependences of the correction
amplitudes in Eqs.~ (\ref{eq18}) and (\ref{eq21}), an ordering
scheme exists for the functions  $\hat \phi_q(t)$ in Fig.
\ref{beta}. Their vertical order before and after crossing the
plateau needs to coincide; this is obeyed in Fig. \ref{beta}.

\begin{figure}[H]
\begin{center}
\includegraphics[width=\columnwidth]{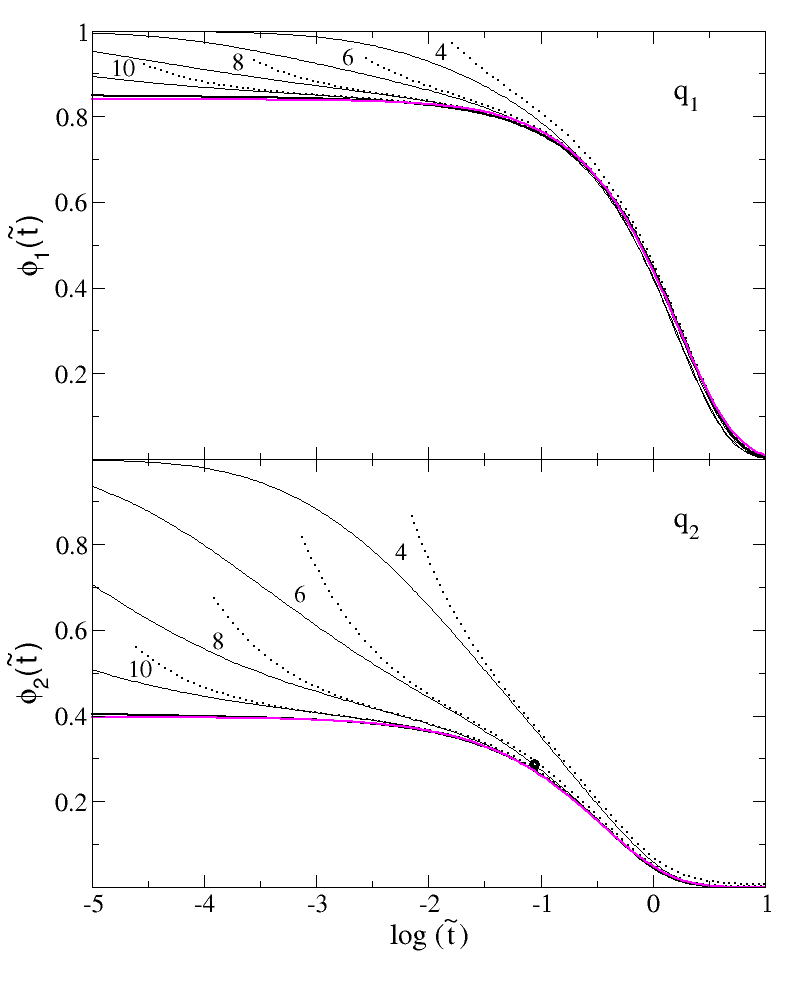}
\caption{\add{(Color online)} Correlators $\phi_q(\tilde t)$ for  different
$\epsilon=-10^{-n/3}$, n=4,6,8,10 as function of rescaled time
$\tilde t=t/t_\sigma^\prime$ (solid lines). The thick solid line is
the $\alpha$-master function $\tilde \phi_q(\tilde t)$. The dotted
lines are the short time parts of leading-plus-next-to-leading
approximation for the $\alpha$-correlators $\tilde \phi_q(\tilde
t)+h_q B_1\sigma \tilde t^{-b}$ according to Eq.~(\ref{eq21z}) with
$B_1=0.5/(\Gamma(1-b)\, \Gamma(1+b)-\lambda)=0.374$. Light (magenta) curves give the Kohlrausch laws fitted to the $\tilde \phi_q(\tilde t)$ in the range $\log_{10}(\tilde t)=[-3.86, 2.14]$; the parameters are given in Fig.~\ref{kohlrausch}.} \label{alpha}
\end{center}
\end{figure}

The second scaling-law regime (for $\varepsilon < 0)$ is defined by
the rescaled time $\tilde{t}=t/t'_\sigma$, where $t'_\sigma$ often
is called $\alpha$-relaxation time and denoted by $\tau$. For
$\tilde{t}={\cal O}(1)$, the $t$-and $\varepsilon$-dependence of
$\phi_q$ is given by the  $\alpha$-master function $\tilde{\phi}_q
(\tilde{t})$ \cite{2}. The validity of this second scaling-law
(Eq.~(\ref{eq21z})) is presented in Figure \ref{alpha}.
Approaching $\varepsilon=0$ from below a collapse onto a
$q$-dependent master function $\tilde{\phi}_q$ occurs.

For the test of the superposition principle, the $\alpha$-relaxation
time was computed  using the power-law Eq.~(\ref{eq20b}) and $t_0$
from Fig.~8. At $q_1$, the $\alpha$-relaxation time obviously
deviates early from the asymptotic power law, since there are
intersections of the rescaled correlators. Nevertheless, the range
of validity of the $\alpha$-scaling law in Fig.~12 far exceeds the
one of the $\beta$-scaling law tested in  Fig.~11. This originates
from the dependence of the leading corrections on the separation
parameter $\epsilon$. While the corrections to the $\beta$-process
are smaller by a factor $\sqrt{\epsilon}$ only, the relative
corrections to the $\alpha$-superposition principle start out in
order $\epsilon$. For example at $q_2$, the $\alpha$ master function
$\tilde\phi_{q_2}(\tilde t)$ describes 68\% of the decay of the
final relaxation better than on a 5\% error level at the separation
$\epsilon=-0.01$ (see the circle in Fig.~\ref{alpha}), while in
Fig.~\ref{beta} for the test of the $\beta$-scaling law smaller
$\epsilon$ are required.

The shape of the $\alpha$-relaxation process, viz.~its master
functions $\tilde\phi_{q}(\tilde t)$, often is described by a
Kohlrausch law
\begin{equation}
\tilde\phi_q(\tilde t) \approx A_q \; \exp{\{-\left(\frac{\tilde t}{\tilde\tau_q}\right)^{\beta_q}\}}
\label{kohl}
\end{equation}
where a possible dependence of the parameters on wavevector $q$ is
taken into account. The von Schweidler  expansion of the
$\alpha$-process, Eq.~(\ref{eq21}), immediately shows that  the
$\tilde\phi_{q}(\tilde t)$ exhibit stretching, viz.~do not decay via
simple exponential relaxation, but result from a broad distribution
of relaxation times. Numerical solutions of the MCT equations in
$d=3$ have shown that, e.g. for hard spheres,  the Kohlrausch law
provides a good overall fit to the master functions for all
wavevectors albeit with noticeable deviations at short rescaled
times, and with $q$-dependent parameters \cite{fhl}. Interestingly
least square fits of Eq.~(\ref{kohl}) to the (final part of the)
$\alpha$-master functions for $\tilde t>10^{-3.86}$, yield
remarkably close agreement of the $d=2$ curves with Kohlrausch laws.
This is shown for two wavevectors in Fig.~12, and can be learned
from the fit parameters presented in Fig.~13.
\begin{figure}[H]
\begin{center}
\includegraphics[width=\columnwidth]{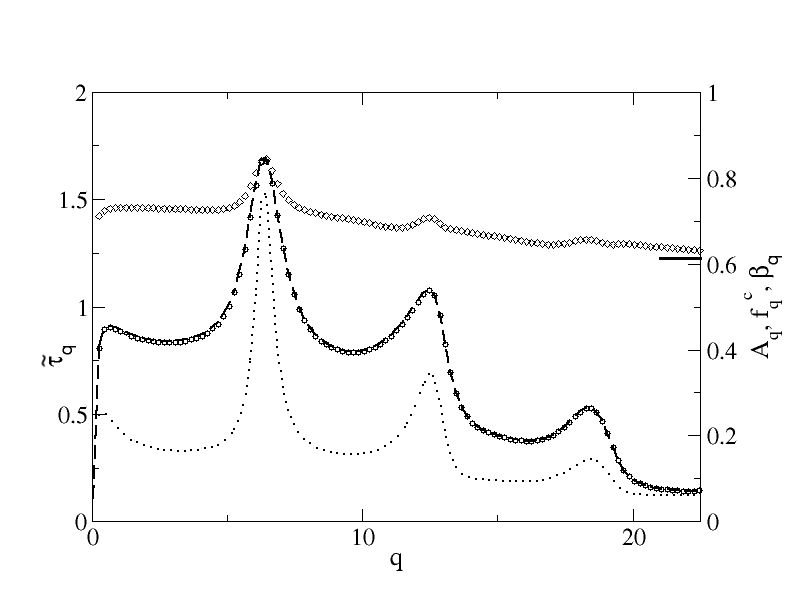}
\caption{Kohlrausch fit parameters of a least square fit of
Eq.~(\ref{kohl}) to the MCT $\alpha$-master functions
$\tilde\phi_q(\tilde t)$ on logarithmic time axis in the interval
$\log_{10}(\tilde t)=[-3.86, 2.14]$. $A_q$ (circles) is quite close
to $f_q^c$ (thick dashed). The Kohlrausch exponent $\beta_q$
(diamonds) converges to the von Schweidler exponent $b=0.613$ (thick
line) for high q values. The times $\tilde \tau_q$ are shown by a
dotted line.} \label{kohlrausch}
\end{center}
\end{figure}
The amplitude $A_q$ of the Kohlrausch law closely follows the
critical NEP, which gives the (true) amplitude of the
$\alpha$-process, and the Kohlrausch stretching exponent $\beta_q$
varies little with wavevectors and is quite close to its large-$q$
limit given by the von Schweidler exponent $b$. Within MCT, the
Kohlrausch law can be  derived as limiting law when an increasing
number of correlators, i.~e.~ correlators with a large range of
$q$-values, contributes to the memory kernels \cite{fu}. This arises
for large wavevectors, then $\beta_q=b$ holds, and the
Kohlrausch-relaxation time depends on wavevector as $\tilde
\tau_q\sim q^{-1/b}$. In the opposite limit, when only one
correlator contributes to the memory function, the schematic ${\cal
F}_2$ model is obtained, where the $\alpha$-process is exponential
\cite{2}. The latter description obviously best applies to the
relaxation of the correlator $\tilde\phi_{q_1}(\tilde t)$ at the
position $q_1$ of the primary peak in the structure factor. It
corresponds to the motion of particles connected to their average
separation, which in MCT is predominantly coupled back to itself.
Guided by these two limits, we speculate that the Kohlrausch law
provides rather good fits to the $\alpha$-process in $d=2$ because
the spread of Kohlrausch exponents between $b=0.61\le \beta_q< 1$ is
smaller in $d=2$ than it is in $d=3$.

\begin{figure}[H]
\begin{center}
\includegraphics[width=1.1\columnwidth]{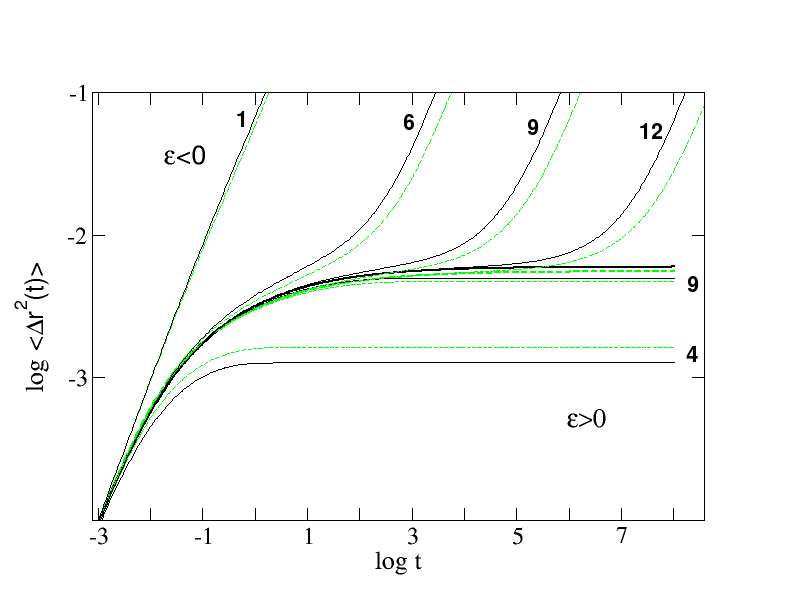}
\caption{\add{(Color online)} Scaled mean squared displacement $\Delta r^2 (t)= \delta
r^2 (t)/(2d)$ in units of $D^2$ for disks in $d=2$ (black solid) and
spheres in $d=3$ (green dashed) for different
$\epsilon=\pm10^{-n/3}$ as labeled. The thick lines are the critical
curves.} \label{MSD23}
\end{center}
\end{figure}
Finally, we have calculated the mean squared displacement $\Delta
r^2 (t)= \delta r^2 (t)/(2d)$, weighted with $1/(2d)$. Figure 14
presents the results obtained from the solution of Eq.~(\ref{eq23})
for $d=2$ and $d=3$ \cite{22}. For the liquid phase the increase of
$\varphi$ leads to the formation of a plateau which has its origin
in the cage effect, independent on $d=2$ or $d=3$. The dynamical
behavior for $t$ small is governed by the short time diffusion
constant $D^{(s)}_0$ and that for $t \rightarrow \infty$ by
$D(\varphi) \sim (t'_\sigma (\varphi))^{-1} \sim
(\varphi_c-\varphi)^\gamma$, the long time diffusivity. At
$\varphi_c$, i.e. for $\varepsilon=0$ a transition occurs where the
cage has an infinite life time such that $D(\varphi)=0$ for $\varphi
\geq \varphi_c$. Consequently the particle becomes localized with
finite localization length $r_s$ (cf.~Eq.~(\ref{eq26})). It is
interesting that $r_s^{d=2} \cong 0.077$ $D$ and $r_s^{d=3} \cong
0.075$ $D$ are almost the same, as can be seen from Figure 14
(as well as from Fig.~\ref{Nep}).

\begin{figure}[H]
\begin{center}
\includegraphics[width=\columnwidth]{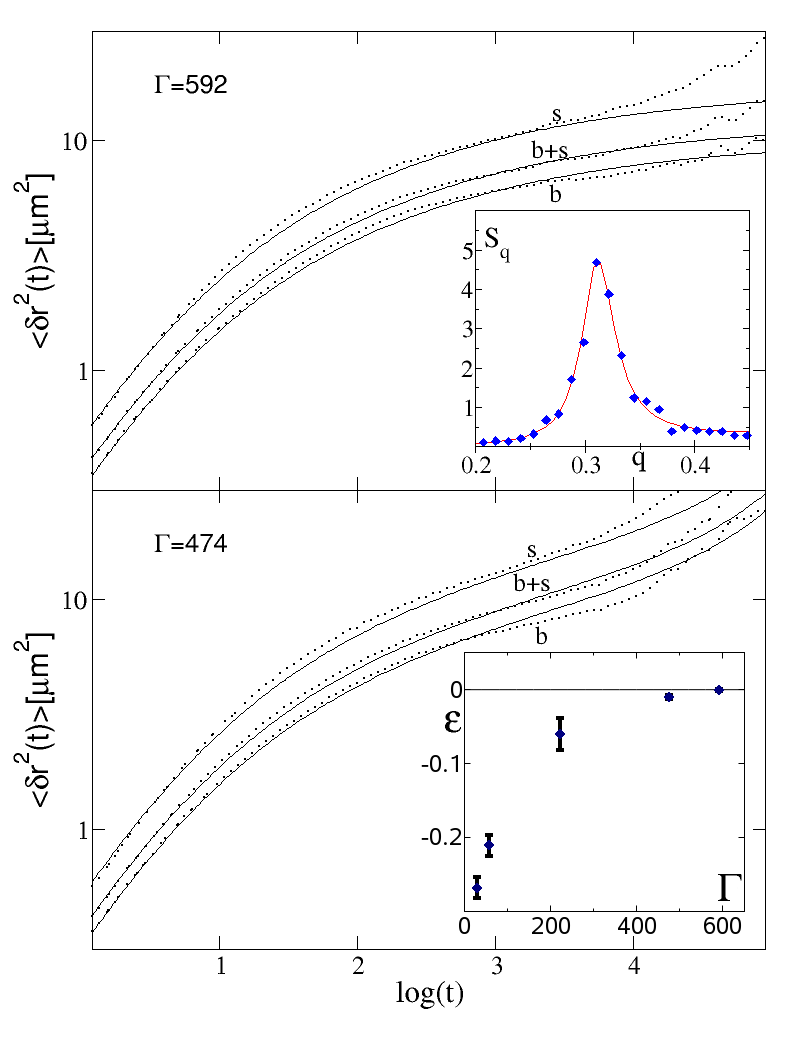}
\caption{Comparison of experimental data (dotted) from \cite{ebertetal} with theoretical curves (solid). Since the
experimental system is a binary mixture three MSD curves are
measured. Curve b is measured  considering only big particles, s
only small particles and b+s takes all particles into account
without discerning big or small. The top panel is at $\Gamma=592$,
$\epsilon=-10^{-3}$, the bottom one at $\Gamma=474$,
$\epsilon=-10^{-2}$. The data at $\Gamma=592$ can in the given
region also be fitted by a critical curve and could thus correspond
to a state in the glass. At $\Gamma=592$, the unit of time is fitted as $\tau^{\rm
BD}= 1/470 $s,  and the unit of length is fitted as (420,700,500)$(\mu m)^2$ for the big (curve b), the small (curve s), and all particles (curve b+s). This leads to a short time diffusion
coefficient $D_0\approx 0.9 e^{-13}m^2/s$  and an effective diameter of $D_{\rm eff}^{\rm big}\approx 20.5 \mu m$ for the big particles. The lower inset shows the fitted separation parameter $\varepsilon$ versus $\Gamma$. The upper inset shows a comparison of the (partial) structure factor of the big particles at $\Gamma=592$ with the one of the hard disk system at $\varepsilon=0$; fortuitously, deviations only appear at larger $q$.} \label{MSDexp}
\end{center}
\end{figure}

Although our results were obtained for monodisperse hard discs we
have made a comparison of the mean squared displacement (MSD) $\delta r^2(t)$ with corresponding
experimental results of the binary system studied in Ref. \cite{18}. The control parameter varied in the experimental system is the interaction parameter, or inverse temperature, $\Gamma$. Interaction parameters larger than  $\Gamma_m=60$ are used, the value at which the corresponding monodisperse experimental system forms a crystal. Because we aim for a qualitative test of our MCT results only, like the existence of finite localization length at the transition, we concentrate on the $\epsilon$-insensitve parts of the MSD's. Thus, we arbitrarily choose $\epsilon\approx0$ for the experimental curve at the highest $\Gamma$ available, $\Gamma=592$. The result of this comparison is shown in Figure \ref{MSDexp}.

In the fit a global length- and timescale was obtained by fitting
the $\Gamma=592$ data. The length- and time scales found at this $\Gamma$ are
then used in fits to data for lower $\Gamma$. Quite good matching to
the data is achieved as shown in Figure \ref{MSDexp} for the example
of $\Gamma=474$. Hence the scaling factors for time- and
lengthscales can be taken as constants and independent of parameter
$\Gamma$. With the scales set, the only fitting parameter left is
$\varphi$, which is adjusted by eye. We find the fits obtained
remarkably good, considering the simplification to map the binary
experimental system onto monodisperse hard discs. From the fitted length 
scale of the single particle MSD of the majority particles (viz. the big ones) 
a reasonable value for the effective particle size follows. The position of the primary peak in the hard-disc structure factor closely agrees with the position of the experimentally obtained (partial) structure factor of the big particles; see the upper inset of Fig.~\ref{MSDexp}. We thus conclude that MCT correctly captures the ratio of localization length to average particle distance.

\section{SUMMARY AND CONCLUSIONS}

In a first step we have determined the explicit dependence of the
MCT functions $\mathcal{F}_q [\phi_k]$ and $\mathcal{F}^{(s)}_q
[\phi_k, \phi_k^{(s)}]$ on the spatial dimensionality $d$. This has
also been done in the hydrodynamic limit $q \rightarrow 0$, which
generalizes the well-known result for $d=3$ \cite{1,21,22} to
arbitrary dimensions.

The major motivation of the present contribution has been the
investigation of the existence of a dynamic glass transition in
two-dimensional systems. As model system we have chosen monodisperse
hard discs which might be a reasonable approximation for a system of
polydisperse hard discs, at least on a qualitative level. Taking the
static input quantity $S_q$ from a modified, hypernetted chain
approximation we have found that an ergodic-nonergodic transition
occurs at a critical packing fraction $\varphi_c^{d=2} \cong 0.697$.
At this critical density, both, the fluctuations of the collective
and self density simultaneously freeze into a glassy state, as for
$d=3$. That $\varphi_c^{d=2}$ is about 35\% above $\varphi_c^{d=3}$
might be not surprising, since the packing fraction
$\varphi^{d=2}_{\rm triang}=\pi/\sqrt{12} \cong0.9069$ of the
triangular lattice is also larger then $\varphi^{d=3}_{\rm hcp}$,
the value for the hexagonal closed packed lattice, by about 20\%.
The difference between $\varphi_c^{d=2}$ and $\varphi_c^{d=3}$
becomes even more obvious when scaling is done with the random close
packing values $\varphi^{d=2}_{\rm rcp} \cong 0.82$ and
$\varphi^{d=3}_{\rm rcp} \cong 0.64$ \cite{24}.
$(\varphi_c/\varphi_{\rm rcp})^{d=2} \cong 0.85$ and
$(\varphi_c/\varphi_{\rm rcp})^{d=3} \cong 0.81$ deviate by not more
than 5\%. Consequently, $\varphi^d_{\rm rcp}$ might be a reasonable
scale for $\varphi^d_c$. So far MCT applied to $d=2$ provides an
explanation for the dynamic glass transition observed in Refs.
\cite{10,11,12}. Since MCT overestimates the glass transition it is
not a surprise that $\varphi^{\rm sim}_c \approx 0.80$ is above
$\varphi_c^{d=2} \cong 0.697$, quite similar to $d=3$ \cite{3}.

For all the investigated quantities and properties connected to the
local 'cage motion' on an intermediate time window, like the
critical nonergodicity parameter, critical amplitudes, the exponent
parameter, the amplitudes $K_q$ and $\bar{K}_q$ of the
next-to-leading order corrections,  etc. we have found a {\it weak}
dependence on dimensionality only, comparing $d=2$ and $d=3$. The
largest change between $d=2$ and $d=3$ occurs for the coherent NEP
at higher wavevectors. For the two-dimensional case, the
$q$-dependence is more abrupt. Otherwise, the  similarity also holds
for the localization length $r_s$ which differ by less than 3\%.
Accordingly, the Lindemann criterium applied to the melting of the
glass phase is almost $d$-independent, at least for $d=2$ and $d=3$.
For longer times, we found the '(time-temperature-) superposition
principle' of the $\alpha$-process; the correlators collapse
onto a non-exponential master function. Interestingly, we observed
for hard discs that the functional form of the $\alpha$-relaxation
closely resembles a Kohlrausch law. This holds better in $d=2$ than
for hard spheres in $d=3$. While the prediction of a superposition
principle for the final decay is guaranteed by the general structure
of the MCT equations, the shape of the relaxation process  provides
information on the local particle rearrangements. Apparently, in the
lower dimension the memory kernels arise from a large number of
contributions so that the Kohlrausch law as a limiting law of large
numbers provides a better approximation to the cooperative motion
during the $\alpha$-decay in $d=2$ than in $d=3$. There is another
conclusion we can draw from our results. On a qualitative level, we
have found consistency with the observations made in Refs.
\cite{14,15,16}. Particularly, the stretching found for the
monodisperse Lennard-Jones system \cite{15} and the binary mixture
of soft discs \cite{16} as well as the two-step relaxation process
\cite{16} can be described by MCT in two dimensions.

Of course, more quantitative comparisons are necessary. MCT should
also be worked out for binary hard discs without and with pair
interactions. This will allow to compare MCT results with the
experimental ones \cite{18} in detail. A first attempt has been
done concerning the mean squared displacement. A more or less
satisfactory agreement has been found over about four decades in
time (see Figure 15). Whether the systematic discrepancies between
the theoretical and the experimental result can be attributed to
the different two model systems, i.e.~on one side the monodisperse
hard discs
and on the other the binary hard discs with dipolar repulsion at
rather low densities, is one of the open questions we intend to
study in the future.

\begin{acknowledgments}
We thank U. Gasser, P. Keim, W. van Megen, and Th. Voigtmann for helpful discussions, and 
acknowledge support by the Deutsche Forschungsgemeinschaft in  SFB 513 and IRTG 667.
\end{acknowledgments}

\end{document}